\documentclass[12pt]{article}
\usepackage{amsmath,amssymb,graphicx,mathrsfs,hyperref}

\newcommand{\be}{\begin{equation}}
\newcommand{\ee}{\end{equation}}
\newcommand{\bea}{\begin{eqnarray}}
\newcommand{\eea}{\end{eqnarray}}

\def\({\left(} \def\){\right)}

\begin{document}
%%%%%%%%%%%%%%%%%%%%%%%%%%%%%%%%%%%%%%%%%
\title{\vspace{-1.8in} The Einstein equations for generalized theories of gravity and the thermodynamic relation $\delta Q = T \delta S$ are equivalent }
\author{\large Ram Brustein ${}^{(1)}$, Merav Hadad ${}^{(1,2)}$  \\ \ \\
(1) Department of Physics, Ben-Gurion University, \\
    Beer-Sheva 84105, Israel,   \\
(2)    Department of Natural Sciences, The Open University of Israel \\
  P.O.B. 808, Raanana 43107, Israel \\
    E-mail: ramyb@bgu.ac.il,\ meravha@openu.ac.il }
\date{}
\maketitle
%%% ----------------------------------------------------------------------

\begin{abstract}
%\abstract{
We show that the equations of motion of  generalized theories of gravity  are equivalent to the thermodynamic relation  $\delta Q = T \delta S$.  Our proof relies on extending previous arguments by using a more general definition of the Noether charge entropy. We have thus completed the implementation of Jacobson's proposal to express Einstein's equations as a thermodynamic equation of state.  Additionally, we find that the Noether charge entropy obeys the second law of thermodynamics if the energy momentum tensor obeys the null energy condition.  Our results support the idea
that gravitation on a macroscopic scale is a manifestation of the thermodynamics of the vacuum.
%}
\end{abstract}

%\keywords{Equation of State, field equation, Entropy}

%\preprint{}

%\begin{document}

\newpage

The profound connection between gravitation and thermodynamics was first suggested by the discovery of black hole (BH) entropy \cite{bek} and Hawking radiation \cite{hawk}. Over a decade ago Jacobson proposed \cite{jacobson} to explain this connection by deriving the Einstein's equations from a thermodynamic equation of state using the proportionality relation of entropy and area for all local acceleration horizons. Elizalde and Silva \cite{Elizalde} extended  Jacobson's proof from the simplest Einstein-Hilbert theory of gravity to more general theories which depend on the Ricci scalar, by using the Noether charge entropy \cite{wald1} rather than assuming that the entropy satisfies a fixed theory-independent proportionality relation to the area. These results support the idea
that gravitation on a macroscopic scale is a manifestation of the thermodynamics of the vacuum state of quantum field theory.

If the relation between gravity and thermodynamics is correct, then it should apply to any metric theory of gravity. Perhaps previous demonstrations of such a relationship were accidental, due to the simplicity of the theory?
In order to strengthen the confidence in the idea and show that previous arguments did not result from an accidental relationship we have extended previous proofs to theories of gravity whose Lagrangian depends on the most general gravitational and matter couplings. The key to proving such a relation is to correctly identify the three quantities in the thermodynamic relation $\delta Q = T \delta S$, the heat transfer $\delta Q$, the temperature $T$ and the entropy $\delta S$.

Generalized theories of gravity appear frequently in the context of effective gravity theories of string theory and supergravity. There the   higher-derivative terms originate from integrating out massive modes or from taking into account quantum corrections. The Lagrangian density of such theories can be expressed as a functional of the metric $g_{ab}$, its Riemann tensor $R_{abcd}$ (and its derivatives) and the matter fields (and their derivatives) which are denoted collectively as $\phi$.  As shown in \cite{wald2}, the total Lagrangian density
$\mathscr{L}\left(g_{ab},R_{abcd},\phi\right)$ can be treated as if $g_{ab}$ and $R_{abcd}$ are independent variables
although $R_{abcd}$ is not an independent field. The correct expression for the variation of $R_{abcd}$ is \cite{gorbon}
\begin{eqnarray}
\label{diff rimman}
\frac{\partial\mathscr{L}}{\partial R_{pabq}}\delta R_{pabq}=2\frac{\partial\mathscr{L}}{\partial R_{pabq}}\nabla_p\nabla_q\delta g_{ab}-\frac{\partial\mathscr{L}}{\partial R_{pabc}}R^q_{\ abc}\delta g_{pq}.
\end{eqnarray}
Note that the sign of the last term on the r.h.s. of (\ref{diff rimman}) is different than its sign in \cite{wald2}. 

Since
$
\frac{\partial\mathscr{L}}{\partial R_{pabq}}\delta R_{pabq}$,  $\frac{\partial\mathscr{L}}{\partial R^p_{\ abq}}\delta R^p_{\ abq}$ at fixed metric  and their counterparts with any number of indices of the Riemann tensor either raised or lowered differ only in index positions, one can find the equations of motion in the following way. First, use the freedom of lowering and raising indices to rewrite the Lagrangian in terms of the Riemann tensor and the minimal possible use of the metric tensor. For example, if the Ricci tensor $R_{ab}$ appears in the Lagrangian density, it is expressed as $R_{ab}=R_{a\ bc}^{\ c}$ and if the Ricci scalar appears in the Lagrangian density it is expressed as $R=R^{ab}_{\ \ ab}$.
Then the Einstein equations are 
\begin{eqnarray}
\label{field-eq1}
\sqrt{-g}\left(-\frac{\partial \mathscr{L}}{\partial g^{ab}} - 2 \nabla_p \nabla_q \frac{\partial\mathscr{L}}{\partial R_{pabq}}+\frac{\partial\mathscr{L}}{\partial R_{pqr}^{\ \ \ a}}R_{pqrb}\right)-\frac{1}{2}\sqrt{-g}g_{ab}\mathscr{L}= 0.
\end{eqnarray}

We now wish to express the Lagrangian as a sum of three terms: a matter Lagrangian  $\mathscr{L}_m(g_{ab},\phi)$ which does not depend on the Riemann tensor, a gravity Lagrangian $\mathscr{L}_G\left(R_{abcd},R^a_{\ bcd},\dots\right)$ which depends only on the Riemann tensor (and its derivatives) with any combination of lowered or raised indices  and an interaction Lagrangian which depends on both $\mathscr{L}_{int}\left(g_{ab},\phi, R_{abcd},R^a_{\ bcd},\dots\right)$.
In particular, the fact that $\mathscr{L}_G$ is independent on the metric $g_{ab}$ is, again, due to the fact that
$
\frac{\partial\mathscr{L}}{\partial R_{pabq}}\delta R_{pabq}$,  $\frac{\partial\mathscr{L}}{\partial R^p_{\ abq}}\delta R^p_{\ abq}$ and their counterparts with any number of indices of the Riemann tensor either raised or lowered differ only in index positions.
The final result is $\mathscr{L}=\mathscr{L}_m\left(g_{ab},\phi\right)+
\mathscr{L}_G\left(R_{abcd},R^a_{\ bcd},\dots\right)+\mathscr{L}_{int}\left(g_{ab},\phi,R_{abcd},R^a_{\ bcd},\dots\right)$.
Here the dots stand for the Riemann tensor with all possible combinations of raised and lower indices, which includes all possible contractions among them.

We define the energy momentum tensor as a sum of the matter contribution and the interaction contribution, $T^{ab}=T^{ab}_m+T^{ab}_{int}$ where the matter contribution is defined in the conventional way  $T^{ab}_m=-2/\sqrt{-g}\ \partial\left(\sqrt{-g}\mathscr{L}_m\right)/\partial g_{ab}$ and $T^{ab}_{int}=-2/\sqrt{-g}\ \partial\left(\sqrt{-g}\mathscr{L}_{int}\right)/\partial g_{ab}$. With these definitions the equations of motion~(\ref{field-eq1}) become
\begin{eqnarray}
\label{field-eq2}
T^{ab}=2\left[-2\nabla_p\nabla_q\frac{\partial\mathscr{L}}{\partial R_{pabq}}+\frac{\partial\mathscr{L}}{\partial R_{pqra}}R_{pqr}^{\ \ \ b}\right]-g^{ab}\mathscr{L}_G.
\end{eqnarray}
For later use we note that by using eq.~(\ref{diff rimman}) and the fact that $\mathscr{L}_{int}$ and  $\mathscr{L}_{G}$  are scalars it follows that
\begin{equation}
\label{cons1}
\nabla_aT^{ab}_{int}=2\nabla_a\left[-2\nabla_p\nabla_q\frac{\partial\mathscr{L}_{int}}{\partial R_{pabq}}+\frac{\partial\mathscr{L}_{int}}{\partial R_{pqra}}R_{pqr}^{\ \ \ b}\right],
\end{equation}
\begin{equation}
\label{cons2}
\nabla^b\mathscr{L}_G=2\nabla_a\left[-2\nabla_p\nabla_q\frac{\partial\mathscr{L}_G}{\partial R_{pabq}}+\frac{\partial\mathscr{L}_G}{\partial R_{pqra}}R_{pqr}^{\ \ \ b}\right].
\end{equation}

As a simple example let consider the case of Einstein's gravity and a matter Lagrangian without an interaction Lagrangian
$
\mathscr{L}=\frac{1}{16\pi G}R+\mathscr{L}_m.
$
In this case $\mathscr{L}_{int}=0$ and thus the only contribution to the energy-momentum tensor comes from the matter $T_{ab}=-2/\sqrt{-g}\ \partial\left(\sqrt{-g}\mathscr{L}_m\right)/\partial g^{ab}$. The Ricci scalar can be expressed as  $R=R^{ab}_{\ \ ab}$ which does not depend on the metric. Since $\frac{\partial R}{\partial R_{pabq}}=\frac{1}{2}\left(g^{pb}g^{aq}-g^{pq}g^{ab}\right)$
we get $\nabla_p\nabla_q\frac{\partial\mathscr{L}}{\partial R_{pabq}}=0$ and the equations of motion~(\ref{field-eq2}) become
$
T^{ab}=2\frac{1}{16\pi G}\left[\frac{\partial\mathscr{L}}{\partial R_{pqra}}R_{pqr}^{\ \ \ b}\right]-g^{ab}\mathscr{L}_G.
$
Substituting the explicit expression for the derivative with respect to the Riemann tensor we find
$
T^{ab}=2\frac{1}{16\pi G}\left[\frac{1}{2}\left(g^{pr}g^{qa}-g^{pa}g^{qr}\right)R_{pqr}^{\ \ \ b}-\frac{1}{2}g^{ab}R\right]
$,
which is indeed the well-known Einstein equation
$
8\pi GT^{ab}=R^{ab}-\frac{1}{2}g^{ab}R
$. We have verified that Eq.~(\ref{field-eq2}) agrees with the conventional derivation also for the more complicated cases when   $\mathscr{L}_{int}$ does not vanish and $\mathscr{L}_{G}$ depends in a general way on the Riemann tensor (and its derivatives).

Since our proof of the equivalence between the Einstein's equations and the thermodynamic relation for generalized theories of gravity is based on Jacobson's proof for the Einstein theory \cite{jacobson}, we briefly recall the fundamental assumptions that were first made by Jacobson: that according to Einstein's  equivalence principle any free-falling local observer can describe space-time in the vicinity of her location as flat. She can also choose the local space-like area element perpendicular to her world-line  at a given point $p_0$. In this setting, the past horizon of $p_0$ is called the �local Rindler horizon� at
$p_0$ and one can define an approximate Killing field generating a boost at $p_0$. Since local Rindler horizons are null and act as causal barriers, they have an entropy S. This entropy measures the correlation with degrees of freedom beyond the horizon and is proportional to the area. A local accelerated observer hovering just inside the horizon sees an energy flow across the causal barrier and a local temperature $T$, the �Unruh temperature� \cite{unruh}.

To extend Jacobson's proposal to all metric theories of gravity we need  specific definitions for the entropy and  the temperature of a causal barrier in generalized theories of gravity. While these quantities have not been defined for causal barriers in such theories, they have been precisely defined for BH's,  so we turn to BH thermodynamics to obtain precise definitions.

We begin with the entropy. We assume that the causal barrier entropy in generalized theories of gravity is the Noether charge entropy (NCE) \cite{wald1}. This assumption was first made in this context by Elizalde and Silva  \cite{Elizalde} and is based on the fact that we expect the causal barrier entropy to be proportional to the area, even in cases where the gravity theory is general.  Since we have already shown in \cite{walderea} that for BH's the NCE is equal to a quarter of the horizon area in units  of the effective gravitational coupling, the assumption that causal barrier entropy in generalized theories of gravity is the NCE seems reasonable.

For the definition of the Unruh temperature in generalized theories of gravity we again turn to BH thermodynamics. BH temperature in any theory of gravity is related to the Killing vector field $\chi_a$ by
\begin{equation}
\label{chikappa}
\chi_b\nabla^b\chi_a=\kappa\chi_a,
\end{equation}
$\kappa$ being the surface gravity, related to the temperature by
\begin{equation}
\label{kappaT}
\kappa=2\pi T.
\end{equation}
We assume that the Unruh temperature satisfies a similar relation, with $\kappa$ being the observer's acceleration.

We now use these ideas to express the energy and entropy for causal barriers in generalized theories of gravity.

Recalling the point $p_0$ with its associated local Rindler horizon $\mathscr{H}$, let us take an accelerated observer hovering just inside the horizon. The energy measured by the observer is  $E =\int_\mathscr{H}T_{ab}\tilde{\chi}^a \epsilon^b$  where the integration is over  a short segment of a thin pencil of horizon generators centered
on the one that terminates at $p_0$.  The normalized Killing field   $\tilde{\chi}^a$ is  null on the horizon and normalized to have unit surface gravity, i.e.  $\chi^a=\kappa \tilde{\chi}^a$.  The vector  $\epsilon ^b =\tilde{\chi}^b\Sigma$ is a $(D-1)$ volume form,  $\Sigma$ being the volume element.  For a constant $\epsilon ^b$ the variation of energy is
\begin{eqnarray}
\label{heat1}
 \delta E=\int_\mathscr{H}\chi^c\nabla_c \left(T_{ab}\chi^a\right) \epsilon^b.
\end{eqnarray}
Since $\chi^c\nabla_c \tilde{\chi}^a=\kappa\tilde{\chi}^a=\chi^a$
\begin{eqnarray}
\label{heat2}
 \delta E=\int_\mathscr{H}\chi^c\nabla_c T_{ab}\tilde{\chi}^a \epsilon^b+\int_\mathscr{H}T_{ab}\chi^a \epsilon^b.
\end{eqnarray}
From eq.~(\ref{heat2}) we deduce that there are two different contributions to $\delta E$. The first contribution  $\chi^d\nabla_d T_{ab}$  is related to real flux of energy that crosses the area. This flux is not directly related to the existence of the causal barrier and does not contribute to the causal barrier entropy. Thus, in agreement with  \cite{jacobson},  we deduce that the heat variation $\delta Q$ that is associated with the causal barrier is
\begin{eqnarray}
\label{heat}
\delta Q =\int_\mathscr{H}T_{ab}\chi^a \epsilon^d.
\end{eqnarray}

We assume that the entropy associated with the causal barrier is the NCE \cite{wald1}:
\begin{eqnarray}
\label{wald1eq}
S=-\frac{1}{T}\oint\limits_{\partial\mathscr{H}}\frac{\partial\mathscr{L}}{\partial R_{abcd}}\hat{\epsilon}_{ab}\epsilon_{cd},
\end{eqnarray}
where the integration is over a surface enclosing the volume $\mathscr{H}$. In Eq.~(\ref{wald1eq})\ $\epsilon^{cd}$ is a (D-2) volume form, $\epsilon^{cd}=\hat{\epsilon}^{cd}\bar{\epsilon}$,  $\bar{\epsilon}$ is the area element on a cross section of the horizon and $\hat{\epsilon}^{cd}$ is the bi-normal vector to the area element and $\hat{\epsilon}^{cd}=\nabla^c\tilde{\chi}^d$.

As pointed out in \cite{kang} (Eq.~(8)), any $W^{cd}$  satisfies  $d\left(W^{cd}\epsilon_{cd}\right)=-2\nabla_cW^{cd}\epsilon_d$ \cite{joey}. Integrating over some volume $V$ this becomes $\oint\limits_{\partial V}W^{cd}\epsilon_{cd}=-2\int\limits_{V}\nabla_cW^{cd}\epsilon_d$. Thus,
\begin{eqnarray}
\label{wald2eq}
S=\frac{2}{T}\int\limits_{\mathscr{H}}\nabla_c\left(\frac{\partial\mathscr{L}}{\partial R_{abcd}}\hat{\epsilon}_{ab}\right)\epsilon_d.
\end{eqnarray}
This leads to
\begin{eqnarray}
\label{wald3eq}
S=\frac{2}{T}\int\limits_{\mathscr{H}}\nabla_c\left(\frac{\partial\mathscr{L}}{\partial R_{abcd}}\right)\hat{\epsilon}_{ab}\epsilon_d +\frac{2}{T}\int\limits_{\mathscr{H}}\frac{\partial\mathscr{L}}{\partial R_{abcd}}\nabla_c\hat{\epsilon}_{ab}\epsilon_d,
\end{eqnarray}
and since $\nabla_c\hat{\epsilon}_{ab}=-R_{abci}\tilde{\chi}^i$ we find
\begin{eqnarray}
\label{wald3eq}
S=\frac{2}{T}\int\limits_{\mathscr{H}}\nabla_c\left(\frac{\partial\mathscr{L}}{\partial R_{abcd}}\right)\hat{\epsilon}_{ab}\epsilon_d -\frac{2}{T}\int\limits_{\mathscr{H}}\frac{\partial\mathscr{L}}{\partial R_{abcd}}R_{abci}\tilde{\chi}^i\epsilon_d.
\end{eqnarray}
The last term vanishes since $\tilde{\chi}^b$ vanishes on the horizon and thus eventually the entropy turns to
\begin{eqnarray}
\label{wald4eq}
S=\frac{2}{T}\int\limits_{\mathscr{H}}\nabla_c\left(\frac{\partial\mathscr{L}}{\partial R_{abcd}}\right)\hat{\epsilon}_{ab}\epsilon_d.
\end{eqnarray}

Now that the integral is over the volume $\mathscr{H}$, we can calculate the entropy variation while keeping  $\epsilon_d$  constant, as in the calculation of the energy variation in Eq.~(\ref{heat1}):
\begin{eqnarray}
\label{entropy-var}
\delta S&=&\frac{2}{T}\int\chi_m\nabla^m\left(\nabla_c\frac{\partial\mathscr{L}}{\partial R_{abcd}}\hat{\epsilon}_{ab}\right)\epsilon_d \cr
&=&4\pi \int\tilde{\chi}_m\nabla^m\left(\nabla_c\frac{\partial\mathscr{L}}{\partial R_{abcd}}\hat{\epsilon}_{ab}\right)\epsilon_d.
\end{eqnarray}
Since we have set the volume vector $\epsilon_d$ to a constant, the entropy variation does not depend on variation of the area, as opposed to Jacobson's assumption in \cite{jacobson}.

Having identified all the ingredients in the  thermodynamic relation
\begin{eqnarray}
\label{thermrel}
\delta Q = T \delta S,
\end{eqnarray}
we can proceed to show that it is equivalent to the equations of motion of  generalized theories of gravity Eq.~(\ref{field-eq2}).

In Eqs.~(\ref{kappaT}),(\ref{heat}) and (\ref{entropy-var}) we have defined the quantities that appear in the thermodynamic relation (\ref{thermrel}). Using them we observe that the thermodynamic relation can only be valid if
\begin{eqnarray}
\label{termo}
\int T^{ab}\tilde{\chi}_a\epsilon_b=2\int \tilde{\chi}_m\nabla^m\left(\nabla_c\frac{\partial\mathscr{L}}{\partial R_{abcd}}\hat{\epsilon}_{ab}\right)\epsilon_d.
\end{eqnarray}
We will now show that Eq.~(\ref{termo}) is indeed valid.

Since $\epsilon _d =\tilde{\chi}_d\Sigma$, then
$
\label{termo1}
\int \tilde{\chi}_m\nabla^m\left(\nabla_c\frac{\partial\mathscr{L}}{\partial R_{abcd}}\hat{\epsilon}_{ab}\right)\epsilon_d= \int \tilde{\chi}_d \tilde{\chi}_m\nabla^m\left(\nabla_c\frac{\partial\mathscr{L}}{\partial R_{abcd}}\hat{\epsilon}_{ab}\right)\Sigma.
$
Using $\tilde{\chi}_m=\tilde{\chi}^n\nabla_n\tilde{\chi}_m=\tilde{\chi}^n\hat{\epsilon}_{nm}$ we can express the integrand of the previous expression as
$
\tilde{\chi}_d\tilde{\chi}^n\hat{\epsilon}_{nm}\nabla^m\left(\frac{\nabla_c\partial\mathscr{L}}{\partial R_{abcd}}\hat{\epsilon}_{ab}\right).
$
Anticipating choosing a point on the horizon, since $\nabla_c\hat{\epsilon}_{ab}=-R_{abci}\tilde{\chi}^i$ and $\tilde{\chi}^b$ vanishes on the horizon we find that
$
\hat{\epsilon}_{nm}\nabla^m\nabla_c\left(\frac{\partial\mathscr{L}}{\partial R_{abcd}}\hat{\epsilon}_{ab}\right)=\hat{\epsilon}_{nm}\hat{\epsilon}_{ab} \nabla^m\nabla_c\frac{\partial\mathscr{L}}{\partial R_{abcd}}.
$
Finally, using $\hat{\epsilon}_{nm}\hat{\epsilon}^{ab}=-(\delta^a_{\ n}\delta^b_{\ m}-\delta^b_{\ n}\delta^a_{\ m})$ and $\hat{\epsilon}^{mi}\hat{\epsilon}_{nm}=\delta^i_{\ n}$ we find
\begin{eqnarray}
\label{field-eq-2,5}
\int T^{ab}\tilde{\chi}_a\epsilon_b=\int2\left[-2\nabla_p\nabla_q\frac{\partial\mathscr{L}}{\partial R_{pabq}}\right]\tilde{\chi}_a\epsilon_b.
\end{eqnarray}
To compare the integrands of both sides of the last eq. we must "choose" from the left side only the terms that are symmetric in $a, b$ . i.e. the last eq. holds if and only if
\begin{eqnarray}
\label{field-eq-4}
 T^{ab}=2\left[-2\nabla_p\nabla_q\frac{\partial\mathscr{L}}{\partial R_{p(ab)q}}\right]+g^{ab}f
\end{eqnarray}
where $R_{p(ab)q}$ is symmetric in $(a, b)$.
Since
\begin{eqnarray}
\label{field-eq-5}
 -2\nabla_p\nabla_q\frac{\partial\mathscr{L}}{\partial R_{p(ab)q}}=-2\nabla_p\nabla_q\frac{\partial\mathscr{L}}{\partial R_{pabq}}+\frac{\partial\mathscr{L}}{\partial R_{pqra}}R_{pqr}^{\ \ \ b}
\end{eqnarray}
 and we find that Eq.~(\ref{termo}) holds if and only if
\begin{eqnarray}
\label{field-eq-3}
T^{ab}=2\left[-2\nabla_p\nabla_q\frac{\partial\mathscr{L}}{\partial R_{pabq}}+\frac{\partial\mathscr{L}}{\partial R_{pqra}}R_{pqr}^{\ \ \ b}\right]+g^{ab}f,
\end{eqnarray}
The  freedom in adding the term $g^{ab}f$ on the r.h.s. of Eq.~(\ref{field-eq-3}) exists because on the horizon $\tilde{\chi}^b$ is null.
The conservation of energy and momentum fixes this freedom up to a constant. Taking the divergence of Eq.~(\ref{field-eq-3}), using the conservation of the matter energy-momentum tensor  and comparing with Eqs.~(\ref{cons1}) and (\ref{cons2}) we observe that  $\nabla^b f=-\nabla^b \mathscr{L}_G$. Thus $f=-\mathscr{L}_G+\Lambda$ (for some constant $\Lambda$) and upon substituting this into Eq.~(\ref{field-eq-3}) it becomes identical to the equations of motion  Eq.~(\ref{field-eq2}).

Turning to the second law of thermodynamics we obtain an additional interesting result.
Equation~(\ref{termo}) can be expressed as
\begin{eqnarray}
\label{termo-s}
\delta S=\frac{1}{2\pi}\int T_{ab}\tilde{\chi}^a\tilde{\chi}^b\Sigma.
\end{eqnarray}
Taking a limit that the volume of integration becomes very small so we can evaluate the integrand as if it were on the horizon we observe that $\delta S\geq 0$ if  $T_{ab}\tilde{\chi}^a\tilde{\chi}^b\geq 0$.
Recall that the energy-momentum tensor satisfies the null energy condition if $T_{ab}X^a X^b\geq 0$ for all null vectors $X^a$. As we just argued since $\tilde{\chi}^a$ is null on the horizon, if the energy-momentum tensor does satisfy  the null energy condition then  the NCE satisfies the second law of thermodynamics:
\begin{eqnarray}
\label{second1}
\delta S\geq 0.
\end{eqnarray}

We may  speculate on the relevance of our results to the issue of the origin of BH entropy. We have assumed that the causal barrier entropy behaves in a similar way to  BH entropy. Since causal barrier entropy is associated with the entanglement with degrees of freedom hidden behind it, we may turn the logic around and speculate that BH entropy also results from entanglement with hidden degrees of freedom. Further investigation may provide a clearer understanding of the suggestive relation between Noether charge entropy and entanglement entropy.

In conclusion, we have shown the equivalence of the equations of motion and thermodynamics for generalized theories of gravity and that the NCE satisfies the second law when the relevant energy conditions are met.

\ \\

{\bf Acknowledgments:}
We thank Dan Gorbonos and Judy Kupferman for discussions and Joey Medved for  comments on the manuscript.
The research of RB and MH was supported by The Israel Science Foundation grant no 470/06. MH research was supported by The Open University of Israel's Research Fund.

\end{document}